\documentstyle[preprint,revtex]{aps}

\begin{document}
\thispagestyle{empty}

% TITLE PAGE

\begin{center}
\hfill{IP-ASTP-08-95}\\
\hfill{April, 1995}

\vspace{1 cm}

\begin{title}
Gravitational Waves in Bianchi Type-I Universes I:
The Classical Theory
\end{title}
\vspace{0.5 cm}

\author{H.~T.~Cho}
\vspace{0.10 cm}

\begin{instit}
Department of Physics, Tamkang University\\ Tamsui, Taipei, Taiwan 251, R.O.C.
\end{instit}
\vspace{0.10 cm}
and
\vspace{0.10 cm}
\author{A.~D.~Speliotopoulos}
\vspace{0.10 cm}
\begin{instit}
Institute of Physics, Academia Sinica\\ Taipei, Taiwan 115, R.O.C.
\end{instit}
\end{center}

\begin{abstract}

\begin{center}
{\bf Abstract}
\end{center}

\noindent The propagation of classical gravitational waves in
Bianchi Type-I universes is studied. We find that gravitational waves
in Bianchi Type-I universes are not equivalent to two minimally
coupled massless scalar fields as it is for the Robertson-Walker
universe. Due to its tensorial nature, the
gravitational wave is much more sensitive to the anisotropy of the
spacetime than the scalar field is and it gains an effective mass
term. Moreover, we find a coupling between the two
polarization states of the gravitational wave which is also not
present in the Robertson-Walker universe.

\vspace{0.25cm}

\end{abstract}
\newpage

\noindent{\bf \S 1. Introduction and Summary}

A Bianchi Type-I (B-I) universe, being the staightforward
generalization of the flat Robertson-Walker (RW) universe, is one of
the simplest models of an anisotropic universe. Unlike a RW
universe which has the same scale factor for each of the three
spatial directions, a B-I universe has a different scale factor in
each direction, thereby introducing an anisotropy to the system.
It moreover has the agreeable property that near the singularity it
behaves like a Kasner universe even in the presence of matter and
consequently falls within the general (classical) analysis of the
singularity given in $\cite{BLK}$. And in an universe filled with
matter for which $p=\gamma\rho$, $\gamma<1$, it has been shown
$\cite{Jacobs}$ that any initial anisotropy in a B-I universe
quickly dies away and a B-I universe will eventually evolve into a RW
universe. Since the present day universe is surprisingly isotropic,
this feature of the B-I universe makes it a prime candidate for
studying the possible effects of an anisotropy in the early universe
on present day observations.

Curiously, in light of the importance of B-I cosmologies the general
behavior of gravitational waves (GW) in a B-I universe has not been
fully analysed. The
propagation of a specific GW in a B-I universe has been studied
before in $\cite{Hu}$. This analysis, however, was only
done for a single wave propagating along the symmetry axis of the
$2/3$, $2/3$, $-1/3$ axial symmetric Kasner spacetime. More
recently GW in B-I universes were studied by Miedema and van Leeuwen
$\cite{Mie}$ within the context of a general perturbative analysis of
the B-I universe. They found, however, certain subtleties with gauge
freedom and gauge fixing which we have not encountered. In
particular, they found that an initially transverse GW will, as it
evolves with time, become longitudinal. This we did not find and we
believe that their result is due to incorrect gauge fixing. We shall
comment further on this after our analysis.

The purpose of this paper, therefore, is to begin the study of
GW's in general B-I spacetimes. The approach we shall take follows
that given in $\cite{Ford}$ which analyses the propagation of GW in
the RW universe. There are, however, certain subtleties in the
propagation of GW in B-I universes which do not occur in RW universes
that makes this analysis much more difficult. These involve, in part, the
choice of the appropriate gauge for the GW as well as the definition
of the polarization tensors for the wave. We find that the usual
transverse-traceless and synchronous gauge conditions that are valid
in the RW universe are inconsistant in the B-I universe. As it is the
gauge conditions which determine the properties of the polarization
tensor, modifications to the gauge conditions were needed to produce
a polarization tensor which has the usual properties expected of a
GW. (Indeed, if we take the usual transverse and synchronous gauges
we would end-up with a GW which is not traceless.) Because, however,
the usual gauge choice is no longer valid in a B-I universe, additional
gauge dependent terms are introduced into the lagrangian which will,
in general, cause a coupling between the two polarization states of
the GW.

The second subtlety we have encountered in defining a GW is in the
definition of the polarization tensors themselves even after a gauge
choice has been made. As usual, the polarization tensor can be expressed in
terms of the appropriate tensor product of two polarization vectors
which are transverse to the propagation direction of the GW. There
is, however, a certain amount of freedom in the choice of these
polarization vectors even after requiring that they be
orthonormal to one another. Namely, one can always do a rotation of
the polarization vectors, which can be time dependent, in the plane
perpendicular to the propagation direction of the GW and the
resultant vectors would still be a valid choice of polarization
vectors for the GW. While this rotational freedom does not play a
role in the isotropic RW universe, the B-I universe is anisotropic
and we have found that an arbitrary choice of
polarization vectors will cause a coupling between the two
polarization vectors. This coupling is fictitious, however, and
vanishes once an appropriate choice of polarization vectors has been made.
In fact something quite similar happens even with GW propagating in RW
universes if we choose polarization vectors which are not orthogonal
to one another. This introduces a coupling between the two
polarization states which vanishes under an appropriate rotation of
polarization vectors.

In general, we find the propagation of GW in B-I universes to be very
much different than in RW universes. In the RW universe the two
polarization states of the GW decouples from one another and the
lagrangian for GW is equivalent to two minimally coupled massless scalar
fields. Neither is true for GW in B-I universes. Due its tensorial
nature, the GW is much more sensitive to the anisotropy of the B-I
universe than a scalar field is and it gains an effective, time
dependent {\it negative\/} masslike term. Moreover, the two
polarizations of the GW now coupled to one another. As this
coupling comes in part from the gauge dependent piece of the
lagrangian, its physical relevance is questionable. In particular, it
is {\it possible\/} that with a more clever choice of gauge this term
will disappear. After our analysis we shall present arguments for the
physical relevance of the coupling terms.

The rest of the paper is divided into five parts. In $\bf \S 2$ we
shall review the basic properties of a B-I cosmology paying
particular attention to the Kasner and Zel'dovich universes. In $\bf
\S 3$ we shall address the question of gauge fixing and the choice of
polarization vectors for the GW. In $\bf \S 4$ we shall derive the
equation of motion for the GW by expanding the GW in plane waves and
writing the action for the GW in momentum space. Then in $\bf \S 5$
we shall solve the evolution equations for the special case of a GW
propagating along an asymmetry axis in a Kasner universe or a
Zel'dovich universe. This is the only case in which closed form
solutions of the equations of motion can be found. Fortunately, they
are the most physically relevant. Concluding remarks can then be
found in $\bf\S 6$. Finally, in the {\bf appendix} we shall repeat
the analysis in $\bf \S 4$ using a different gauge choice for the
gravitational wave and show that coupling terms similar to those
found in $\bf \S 4$ appear even in this gauge choice.

\noindent{\bf \S 2. Review of B-I Cosmology}

In this section we shall present a brief review of B-I cosmologies in
the absence of perturbations. In doing so we shall follow the
notational and sign conventions found in $\cite{BirDav}$. In
particular, one may always choose coordinates for the B-I spacetime
such that the metric has the form
\begin{equation}
ds^2 = dt^2 - \sum_{i=1}^{3}a_i^2(t)(dx^i)^2\>,
\label{e1}
\end{equation}
where as usual Greek indices will run over the four spacetime
directions while Roman indices will run only over the spatial
directions. Although we shall use the summation convention for Greek
indices, for clarity we shall {\it not\/} use it for Roman indices.

It is convenient to define a pseudo-conformal time variable $\eta$
through
\begin{equation}
\frac{d\eta}{dt} = \frac {1}{a(\eta)}\>,
\label{e2}
\end{equation}
where $a\equiv (a_1a_2a_3)^{1/3}$ is the geometric average of the
three scale factors along each direction. Then the only non-vanishing
components of the Riemann tensor are
\begin{equation}
R_{0i,0i} =-\frac{C_j}{2}
	   \left(
	   	d'_i+\frac{d_i^2}{2}-\frac{d_iD}{2}
	   \right)\>,
\label{e3}
\end{equation}
while for $i\ne j$,
\begin{equation}
R_{ij,ij} = \frac{1}{4}\frac{d_id_jC_iC_j}{C}\>,
\label{e4}
\end{equation}
where $C=a^2(\eta)$, $C_j=a^2_j(\eta)$,
\begin{equation}
d_j \equiv \frac{C'_j}{C_j}\>, \qquad D \equiv \frac{C'}{C} =
\frac{1}{3} \sum_{i=1}^3 d_i\>,
\label{e5}
\end{equation}
and the prime denotes derivative with respect to $\eta$.
Einstein's equations in the presence of matter are then
\begin{eqnarray}
\frac{3}{2}D' + 6Q &=& -4\pi C(\rho+3p)\>,
\nonumber \\
d_j'+d_jD &=& 8\pi C(\rho-p)\>,
\label{e6}
\end{eqnarray}
where
\begin{equation}
Q \equiv \frac{1}{72}\sum^3_{i>j} (d_i-d_j)^2\>,
\label{e7}
\end{equation}
and
\begin{equation}
\left(
	\frac{1}{a}
	\frac{da}{d\eta}
\right)^2
	= \frac{8\pi a^2}{3}\rho + Q\>.
\label{e8}
\end{equation}
This, aside from the term dependent on $Q$, is identical to the RW
case. $Q$ is thus the physical measure of the anisotropy of the spacetime
and is called the anisotropy factor. Taking then the difference
between two different directions in eq.~$(\ref{e6})$, we find that
\begin{equation}
d_i-d_j = (d_i^0-d_j^0)/a ^2\>,
\label{e9}
\end{equation}
where $d^0_i$ is the value of $d_i$ at some initial time $\eta_0$ for
all $\rho$ and $p$ and we have chosen the overall scale such that
$a(\eta_0)=1$. Consequently, $Q = Q_0/a^4$ where $Q_0=Q(\eta_0)$.

Eq.~$(\ref{e6})$ has been solved in the presence of various types of
matter $\cite{Jacobs}$. We shall, however, only be concerned with two
special cases in this paper: the Kasner universe, which is free of
matter, and the Zel'dovich universe where $\rho = p$. In both cases,
\begin{equation}
a^2 = y/y_0\>, \qquad a_j^2  = {a_0}_j^2(y/y_0)^{3p_j}\>,
\label{e10}
\end{equation}
where $y = 2\eta \sqrt{Q_0}$ for the Kasner universe while $y = 2
\eta\sqrt{8\pi\rho_0/3 + Q_0}$ for the Zel'dovich universe.
$\rho_0$ is the energy density of the universe at $\eta_0$ while
${a_0}_j =a_j(0)$ are the initial scale factors in the various
directions which, without loss of generality, can be set to unity.

$p_j$ are parameters of the B-I spacetime which measure the {\it
relative\/} anisotropy between any two asymmetry axis. As they must
satisfy the constraints
\begin{equation}
1= \sum^3_{i=1} p_i\>, \qquad 1 = \sum^3_{i=1} p_i^2\>,
\label{e11}
\end{equation}
out of the three parameters, only one is arbitrary. Since
eq.~$(\ref{e11})$ describes the intersection of a sphere with a plane
in the parameter space $(p_1, p_2, p_3)$, we may parametrize the
allowed values of $p_j$ by an angle on the unit circle $\cite{com1}$.
One particular choice of parametrization is
\begin{eqnarray}
p_1 &=& \frac{1}{3} \left(1+\cos\theta+\sqrt{3}\sin\theta\right)\>,
\nonumber \\
p_2 &=& \frac{1}{3} \left(1+\cos\theta-\sqrt{3}\sin\theta\right)\>,
\nonumber \\
p_3 &=& \frac{1}{3} \left(1-2\cos\theta\right)\>,
\label{e12}
\end{eqnarray}
Although a prior\'\i{} $\theta$ ranges over the unit circle, note
that the labeling of each $p_j$ is quite arbitrary. Thus the unit circle
can be divided into six equal parts each of which span $60^0$, and
the choice of $p_j$ is unique within each section separately. Notice
that when $\theta=0$, $p_1=p_2=2/3$ while $p_3=-1/3$. This is the
spacetime considered by Hu in $\cite{Hu}$. When $\theta=\pi/3$, on
the other hand, $p_1=1$ while $p_2=p_3=0$. It can be shown that this
spacetime is equivalent to the Minkowski spacetime up to a coordinate
transformation.

\noindent{\bf \S 3. Gauge Conditions and Polarization States}

In this section we address the problem of gauge fixing for the GW and
the subsequent definition of polarization states. As usual, we
consider the GW as a perturbation off the background metric by
writing $g_{\mu\nu} = g^0_{\mu\nu} + h_{\mu\nu}$, where
$g^0_{\mu\nu}$ is the unperturbed B-I metric given in
eq.~$(\ref{e1})$ while $h_{\mu\nu}$ is the perturbation which
has a gauge freedom to be taken care of. The usual gauge
choice, which works for both flat Minkowski and RW spacetimes, is
\begin{equation}
\nabla^\mu h_{\mu\nu} = 0\qquad,\qquad u^\mu h_{\mu\nu} = 0
\qquad,\qquad  h^\mu_\mu= 0\>,
\label{e13}
\end{equation}
and are called the transverse, synchronous and traceless
conditions, respectively. $u_\mu$ is a timelike Killing vector which,
without lose of generality, can be taken to be $(1,0,0,0)$. In
the RW spacetime, the transverse and synchronous condition actually
implies that the GW is traceless and this condition is
redundant. Unfortunately, for a B-I universe this is no longer true.

Let us express $h^{(s)}_{\mu\nu}$ as a plane wave
\begin{equation}
h^{(s)}_{\mu\nu} = \varpi^{(s)}_{\mu\nu}(\eta) h_{\vec q s}(\eta)
e^{i\vec q\cdot \vec x}\>,
\label{e14}
\end{equation}
where $\vec q \cdot \vec x = \sum_j q_j x^j$ and $s$ labels the
polarization state. Because $g^0_{\mu\nu}$ is a function of $\eta$
only, we can always make this choice. Our convention is that the
$q_j$ are independent of $\eta$ while $q^j = g^{jj}q_j$.
$\varpi^{(s)}_{\mu\nu}$ is the polarization tensor for the GW and
must, by definition, be independent of the amplitude $h_{\vec q
s}(\eta)$ of the GW. The usual transversality condition then gives
\begin{equation}
0=g^{\mu\nu}\partial_\mu \varpi^{(s)}_{\nu0} +
{\varpi^{(s)}}^\mu_0\partial_\mu\log h_{\vec qs} + i \sum_j
q^j\varpi^{(s)}_{j0} +\sum_j
\frac{a_j'}{a_j}{\varpi^{(s)}}_0^0-\sum_j
\frac{a_j'}{a_j}{\varpi^{(s)}}_j^j \>,
\label{e15}
\end{equation}
as well as
\begin{equation}
0=g^{\mu\nu}\partial_\mu \varpi^{(s)}_{\nu k} +
{\varpi^{(s)}}^\mu_k\partial_\mu\log h_{\vec qs} + i \sum_j
q^j\varpi^{(s)}_{jk}+2\frac{a'}{a}{\varpi^{(s)}}^0_k\>.
\label{e16}
\end{equation}
{}From this, we see that for the polarization tensor to be independent
of $h_{\vec qs}$, we {\it must\/} choose the synchronous gauge. After
doing so, the transversality condition reduces to
\begin{equation}
0= \sum_j \frac{a_j'}{a_j} {\varpi^{(s)}}_j^j\qquad,\qquad 0 = \sum_j q^j
\varpi^{(s)}_{j\nu}\>.
\label{e17}
\end{equation}
and becomes the standard transverse-traceless condition in a RW
universe. For a B-I universe, on the other hand, we see that the GW
is no longer traceless if we use the usual transverse and
synchonous gauges. The usual gauge choices eq.~$(\ref{e13})$ are no
longer self-consistant in a B-I universe and must be modified.

There are two straightforward modifications that we can make. The
first is to use the usual transverse and synchronous gauges and live with
a GW which is not traceless. This, however, introduces a great deal
of complexity to the problem and we shall not do so (see {\bf
appendix}). The second is to require that the GW be traceless, but
to modify the transversality condition. Namely, to choose
\begin{equation}
0 = \nabla^\mu h_{\mu j}\qquad,\qquad 0 = h_{0\mu}
\qquad,\qquad 0=h^\mu_\mu \>,
\label{e18}
\end{equation}
which gives the following constraints on the polarization tensor
\begin{equation}
0 = \sum_j q^j\varpi^{(s)}_{jk} \qquad,\qquad 0
=\varpi^{(s)}_{0\nu}\qquad,\qquad 0=\sum_j{\varpi^{(s)}}_j^j\>.
\label{e19}
\end{equation}
Notice that these equations determining $\varpi^{(s)}_{\mu\nu}$ have
precisely the same form as the those in either a RW or Minkowski
spacetime which is the reason we shall work primarily in this gauge.

Instead of the polarization tensor $\varpi^{(s)}_{ij}$, we shall find
it conceptually easier to work with the polarization vectors
$\epsilon^{(s)}_\mu$
with
\begin{equation}
\varpi^{(+)}_{jk} = \epsilon^{(1)}_j\epsilon^{(1)}_k -
\epsilon^{(2)}_j\epsilon^{(2)}_k
\qquad,\qquad
\varpi^{(\times)}_{jk} = \epsilon^{(1)}_j\epsilon^{(2)}_k +
\epsilon^{(2)}_j\epsilon^{(1)}_k
\label{e20}
\end{equation}
as long as
\begin{equation}
0 = \sum_j q^j\epsilon^{(s)}_j \qquad,\qquad \sum_j
\epsilon^{(s)}_j{\epsilon^{(s')}}^j =-\delta_{ss'}\>.
\label{e21}
\end{equation}
Implicit in this definition is the added requirement that the two
polarization states be ``orthonormal'' to one another:
$\varpi^{(s)}_{\mu\nu}{\varpi^{(s')}}^{\mu\nu} = 2\delta^{ss'}$.
Notice, however, that this orthonormality condition is {\it not\/}
required by the gauge choice eq.~$(\ref{e19})$ but is rather made
separately as a minimal requirement for the two polarization states to
decouple from one another. While such a choice is sufficient for a RW
or Minkowski spacetime, we shall see in the next section that
additional requirements are needed for a B-I spacetime.

Eq.~$(\ref{e21})$ does not determine $\epsilon^{(s)}_j$ uniquely,
however. There is still a rotational freedom in the plane
perpendicular to $q_j$. Namely, we can rotate
\begin{eqnarray}
\hat \epsilon^{(1)}_j &=& \epsilon^{(1)}_j \cos\phi + \epsilon^{(2)}_j
\sin\phi \>,
\nonumber \\
\hat \epsilon^{(2)}_j &=& -\epsilon^{(1)}_j \sin\phi + \epsilon^{(2)}_j
\cos\phi \>,
\label{e22}
\end{eqnarray}
and $\hat\epsilon^{(s)}_j$ will still satisfy eq.~$(\ref{e21})$. In
particular, $\phi$ may also be $\eta$-dependent. Moreover, once a
specific choice of polarization vectors has been made
rotational symmetry will be broken. While this breaking of rotational
symmetry is of no consequence in the RW universe, which is isotropic
to begin with, it is of great consequence in the anisotropic B-I
universe. In particular, any specific choice of polarization vectors
will tend to break the expected exchange symmetry $1$-$2$-$3$ among
the labels of the axis of the spacetime. We shall also find
that the equations of motion for the GW will be greatly simplified if
we make a special choice of polarization vectors. For these reasons we
shall work as much as possible with a general set of polarization
vectors and shall delay making a specific choice of polarization
vectors until forced to.

To find a specific representation of $\epsilon^{(s)}_j$ we first define a
local coordinate system using the vierbeins $e_\mu^i$,
\begin{equation}
e_\mu^ie^{\mu j} = -\delta^{ij}\>.
\label{e23}
\end{equation}
Then, taking $e_\mu^1=(0,a_1,0,0)$, $e_\mu^2=(0,0,a_2,0)$,
$e_\mu^3=(0,0,0,a_3)$, one specific choice of polarization vectors is
\begin{eqnarray}
\epsilon^{(1)}_j = &{}&
	\frac{\hat q_1\hat q_3}{\sqrt{\hat q_1^2+\hat q_2^2}} e^1_{j}
	+
	\frac{\hat q_2\hat q_3}{\sqrt{\hat q_1^2+\hat q_2^2}} e^2_{j}
	-
	\sqrt{\hat q_1^2+\hat q_2^2}e^3_{j}\>.
\nonumber \\
\epsilon^{(2)}_j = &{}&
	-\frac{\hat q_2}{\sqrt{\hat q_1^2+\hat q_2^2}} e^1_{j}
	+
	\frac{\hat q_1}{\sqrt{\hat q_1^2+\hat q_2^2}} e^2_{j} \>,
\label{e24}
\end{eqnarray}
where
\begin{equation}
\hat q_j = - \frac{q_j/a_j}{\sqrt{\sum_l q^2_l/a_l^2}}\>.
\label{e25}
\end{equation}
Notice in particular that $\epsilon^{(s)}_j(-\vec q) =
(-1)^{s+1}\epsilon^{(s)}_j(\vec q)$ so that $\varpi^{(+)}_{jk}(-\vec q) =
\varpi^{(+)}_{jk}(\vec q)$ while $\varpi^{(\times)}_{jk}(-\vec q) =
-\varpi^{(\times)}_{jk}(\vec q)$ .

The polarization tensors defined in eq.~$(\ref{e24})$ are for
linearly polarized GW. We shall also consider circularly polarized GW
defined through
\begin{equation}
\varpi^{(L)}_{jk} = \frac{1}{\sqrt{2}}\left(\varpi^{(+)}_{jk} +
		i\varpi^{(\times)}_{jk}\right)
\qquad,\qquad
\varpi^{(R)}_{jk} = \frac{1}{\sqrt{2}}\left(\varpi^{(+)}_{jk} -
		i\varpi^{(\times)}_{jk}\right)\>,
\label{e26}
\end{equation}
with $\varpi^{(L)}(-\vec q) = \varpi^{(R)}(\vec
q)=\overline\varpi^{(L)}(\vec q)$ and $\varpi^{(R)}(-\vec q) =
\varpi^{(L)}(\vec q)=\overline\varpi^{(R)}(\vec q)$
and the bar denotes complex conjugation. We can, of course, also
define circularly polarized polarization vectors in precisely the
same way
\begin{equation}
\epsilon^{(L)}_j = \frac{1}{\sqrt{2}}\left(\epsilon^{(1)}_j +
		i\epsilon^{(2)}_j\right)
\qquad,\qquad
\epsilon^{(R)}_j = \frac{1}{\sqrt{2}}\left(\epsilon^{(1)}_j -
		i\epsilon^{(2)}_j\right)\>,
\label{e27}
\end{equation}
and there is a particularly simple relationship between $\varpi$ and
$\epsilon$,
\begin{equation}
\varpi^{(L)}_{jk} = \sqrt2 \epsilon^{(L)}_j\epsilon^{(L)}_k\qquad,\qquad
\varpi^{(R)}_{jk} = \sqrt2 \epsilon^{(R)}_j\epsilon^{(R)}_k .
\label{e28}
\end{equation}
Finally, we note that under a rotation of polarization vectors as in
eq.~$(\ref{e22})$, $\hat\epsilon^{(L)}_j = e^{-i\phi}\epsilon^{(L)}_j$,
$\hat\epsilon^{(R)}_j = e^{i\phi}\epsilon^{(R)}_j$.

\noindent{\bf \S 4. The Action}

In this section we shall derive the equation of motion for the GW.
The approach we shall follow was first used by Ford and Parker
$\cite{Ford}$ in their analysis of GW in RW universes and involves
the expansion of the GW in plane waves. To lowest
order the action for the GW is
\begin{eqnarray}
I = \frac{1}{4}\int \sqrt{-g}d^4x
    \Bigg(&{}&
    	\nabla_\mu h_\alpha^\beta\nabla^\mu h^\alpha_\beta +
	8\pi(\rho-p)h_\alpha^\beta h^\alpha_\beta
	- 2R_\beta^\mu h^\alpha_\mu h_\alpha^\beta
\nonumber \\
	&{}&
	-2{R^{\mu\beta}}_{\alpha\nu} h_\mu^\nu h^\alpha_\beta
	-2\nabla_\rho h^{\mu\rho}\nabla_\alpha h^\alpha_\mu
	\Bigg)\>,
\label{e29}
\end{eqnarray}
which differs from the one used in $\cite{Ford}$ for GW in RW
spacetimes by the addition of a kinetic term. This is because our
gauge choice eq.~$(\ref{e18})$ is different from the one they used as
in eq.~$(\ref{e13})$. We should also mention that even if we did use the
usual transverse and synchronous gauges, our action will have
additional terms dependent on $h^\mu_\mu$ which no longer vanishes in
this gauge for a B-I universe. Indeed, {\it any\/} choice of gauge
for the GW in a B-I universe will introduce additional terms to the
lagrangian. The form that these additional terms take is
dependent on the choice of gauge one makes, however, and at first
glance the physical relevance of these terms, since they are
seemingly gauge dependent, is questionable. We shall see, however,
that although the explicit form of these terms changes with the
choice of gauge, they all have the same root physical cause. Namely,
they are due to the fact that in a B-I universe one cannot define a
polarization state for the GW which will not change with time.
Consequently, these terms are physically relevant.

It is straightforward to show that due to Einstein's equations, the
second and third terms in eq.~$(\ref{e29})$ cancel one another. We
are then left with three terms to consider
\begin{eqnarray}
I_K &=&\frac{1}{4}\int a^4d^4x
      	\nabla_\mu h_\alpha^\beta\nabla^\mu h^\alpha_\beta\>,
\nonumber \\
I_R &=& -\frac{1}{2}\int a^4d^4x {R^{\mu\beta}}_{\alpha\nu} h_\mu^\nu
	h^\alpha_\beta\>,
\nonumber \\
I_g &=& -\frac{1}{2}\int a^4d^4x \nabla_\rho h^{\mu\rho}\nabla_\alpha
h^\alpha_\mu\>,
\label{e30}
\end{eqnarray}
where $I=I_K + I_R + I_g$. We next expand $h_{\mu\nu}$ in plane waves,
\begin{equation}
h_{\mu\nu} = \sum_{\vec q}\sum_{s=+,\times} \varpi^{(s)}_{\mu\nu}
h_s(\vec q,x)\>,
\label{e31}
\end{equation}
where $h_s(\vec q,x) = h_{\vec q s}e^{i\vec q\cdot \vec x}$ with the
reality condition given by
\begin{equation}
\bar h_{+}(\vec q,x) = h_{+}(-\vec q,x)\qquad,\qquad
\bar h_{\times}(\vec q,x) = -h_{\times}(-\vec q,x)\>.
\label{e32}
\end{equation}

Turning our attention first to $I_R$, we use eq.~$(\ref{e31})$ and
find that
\begin{equation}
I_R = -\frac{1}{2}\sum_{\vec q} \sum_{ss'}\int a^4 d^4 x
T_{ss'}(\epsilon) \bar h_{s'}(\vec q, x) h_{s}(\vec q, x) \>,
\label{e33}
\end{equation}
where
\begin{equation}
T_{ss'}(\epsilon) =
R^{\mu\rho\alpha\nu}\varpi^{(s)}_{\mu\nu}\varpi^{(s')}_{\rho\alpha}
\label{e34}
\end{equation}
In terms of polarization vectors,
\begin{equation}
T_{++}(\epsilon) = T_{\times\times}(\epsilon) = -2\sum_{ijkl}
R^{ijkl}\epsilon^{(1)}_i\epsilon^{(1)}_l\epsilon^{(2)}_j\epsilon^{(2)}_k
\>,
\label{e35}
\end{equation}
while
\begin{equation}
T_{+\times}(\epsilon) = T_{\times +}(\epsilon) = 0\>,
\label{e36}
\end{equation}
due to the anti-symmetry properties of the curvature tensor. Next, using
eq~$(\ref{e4})$,
\begin{equation}
T_{++}(\epsilon) = \frac{2}{a^2}
		   \left\{
		   \left(
		   \sum_i\frac{a_i'}{a_i}\epsilon^{(1)}_i{\epsilon^{(1)}}^i
		   \right)
		   \left(
		   \sum_j\frac{a_j'}{a_j}\epsilon^{(2)}_j{\epsilon^{(2)}}^j
		   \right)
		   -
		   \left(
		   \sum_j\frac{a_j'}{a_j}\epsilon^{(1)}_j{\epsilon^{(2)}}^j
		   \right)^2
		   \right\}\>,
\label{e37}
\end{equation}
where it is important to note that we have not as yet chosen a
specific polarization vector $\epsilon^{(s)}_j$.

We next consider the gauge term which becomes
\begin{eqnarray}
I_g = -\frac{1}{2} \sum_{\vec q}\sum_{ss'} \int a^4 d^4 x
	&{}&
	\Bigg\{
	\varpi^{(s')}_{\mu\rho}{\varpi^{(s)}}^{\mu_\alpha}\nabla^\rho
	\bar h_{s'}\nabla_\alpha h_s
	+
	2\varpi^{(s')}_{\mu\rho}\nabla_\alpha{\varpi^{(s)}}^{\mu\alpha}
	\nabla^\rho \bar h_{s'}h_{s}
	+
\nonumber \\
	&{}&
	\nabla^\rho\varpi^{(s')}_{\mu\rho}\nabla_\alpha
	{\varpi^{(s)}}^{\mu\alpha} \bar h_{s'}h_s
	\Bigg\}\>,
\label{e38}
\end{eqnarray}
for plane waves. The first two terms involve the term
${\varpi^{(s)}}_{\mu\alpha}\nabla^\alpha h_s(\vec q,\eta)$ which
vanishes identically due to eq.~$(\ref{e19})$. The last term does
not, however, and if we define
\begin{equation}
N^2_{ss'} = -\nabla^\rho\varpi^{(s')}_{\mu\rho}
	     \nabla^\alpha{\varpi^{(s)}}^\mu_\alpha
\label{e39}
\end{equation}
we find that
\begin{eqnarray}
N^2_{++} &=&
-\sum_i\left(
	\epsilon^{(1)}_i\nabla^i\epsilon^{(1)}_\mu-
	\epsilon^{(2)}_i\nabla^i \epsilon^{(2)}_\mu
\right)
\sum_j
\left(
	\epsilon^{(1)}_j\nabla^j{\epsilon^{(1)}}^\mu-
	\epsilon^{(2)}_j\nabla^j{\epsilon^{(2)}}^\mu
\right)
\nonumber \\
&=& -\frac{1}{a^2}\left(\sum_j\frac{a_j'}{a_j}{\varpi^{(+)}}_j^j\right)^2
\nonumber \\
N^2_{\times\times} &=&
-\sum_i\left(
	\epsilon^{(1)}_i\nabla^i\epsilon^{(2)}_\mu +
	\epsilon^{(2)}_i\nabla^i \epsilon^{(1)}_\mu
\right)
\sum_j
\left(
	\epsilon^{(1)}_j\nabla^j {\epsilon^{(2)}}^\mu+
	\epsilon^{(2)}_j\nabla^j {\epsilon^{(1)}}^\mu
\right)
\nonumber \\
&=& -\frac{1}{a^2}\left(\sum_j\frac{a_j'}{a_j}{\varpi^{(\times)}}_j^j\right)^2
\nonumber \\
N^2_{+\times} &=&
-\sum_i\left(
	\epsilon^{(1)}_i\nabla^i\epsilon^{(1)}_\mu-
	\epsilon^{(2)}_i\nabla^i\epsilon^{(2)}_\mu
\right)
\sum_j
\left(
	\epsilon^{(1)}_j\nabla^j{\epsilon^{(2)}}^\mu+
	\epsilon^{(2)}_j\nabla^j{\epsilon^{(1)}}^\mu
\right)
\nonumber \\
&=& -\frac{1}{a^2}\left(\sum_i\frac{a_i'}{a_i}{\varpi^{(+)}}_i^i\right)
\left(\sum_j\frac{a_j'}{a_j}{\varpi^{(\times)}}_j^j\right)
\label{e40}
\end{eqnarray}
while $N^2_{+\times}=N^2_{\times+}$ and we have used $\nabla^\mu
\epsilon^{(s)}_\mu=0$. Notice, in particular, that in general
$N^2_{+\times}\ne0$ and this gauge term introduces a coupling
between the two polarization states. In addition, $N_{++}^2 \ne
N_{\times\times}^2$. Notice also that each of the $N^2$ are related
in a simple manner to the trace of the polarization vectors which would
vanish under the usual choice of gauge eq.~$(\ref{e17})$.

Finally, we turn our attention to the kinetic piece, the most
difficult one to work with. Proceeding exactly as before, we find
that
\begin{equation}
I_K = \frac{1}{2}\sum_{\vec q}\int a^4 d^4x \left\{
	\sum_s \nabla_\rho\bar h_s\nabla^\rho h_s
	-
	\sum_{ss'} M^2_{ss'}\bar h_{s'} h_s
	-
	2\sum_{ss'} {D_\mu}_{ss'}\bar h_{s'} \nabla^\mu h_s
	\right\}\>,
\label{e41}
\end{equation}
where
\begin{equation}
{D_\mu}_{ss'} = -\frac{1}{2}
{\varpi^{(s)}}^\alpha_\beta\nabla_\mu{\varpi^{(s')}}^\beta_\alpha
\qquad,\qquad
M^2_{ss'} = -\frac{1}{2} \nabla_\rho\varpi^{(s)}_{\mu\nu}
		   \nabla^\rho {\varpi^{(s')}}^{\mu\nu}\>.
\label{e42}
\end{equation}
This can be reduced to manageable form by using polarization vectors:
\begin{equation}
M^2_{++} = M^2_{\times\times}
	= \sum_s \nabla_\rho\epsilon^{(s)}_\mu
	  \nabla^\rho{\epsilon^{(s)}}^\mu -
	  2(\epsilon^{(1)}_\mu\nabla_\rho{\epsilon^{(2)}}^\mu)
	  (\epsilon^{(1)}_\nu\nabla^\rho{\epsilon^{(2)}}^\nu)\>,
\label{e43}
\end{equation}
where we have used $\epsilon^{(s')}_\mu{\epsilon^{(s)}}^\mu=-\delta_{ss'}$,
while $M_{+\times} \equiv 0$. Similarly,
\begin{equation}
{D_\mu}_{++} = {D_\mu}_{\times\times} =
	\epsilon^{(1)}_\alpha \nabla_\mu{\epsilon^{(1)}}^\alpha
	+
	\epsilon^{(2)}_\alpha \nabla_\mu{\epsilon^{(2)}}^\alpha\>,
\label{e44}
\end{equation}
which vanishes identically while
\begin{equation}
{D_\mu}_{+\times} = -{D_\mu}_{\times +} =
	\epsilon^{(1)}_\alpha \nabla_\mu{\epsilon^{(2)}}^\alpha
	-
	\epsilon^{(2)}_\alpha \nabla_\mu{\epsilon^{(1)}}^\alpha\>.
\label{e45}
\end{equation}
This term will not vanish in general, however, and we see that the
kinetic piece of the lagrangian also introduces a coupling between
the two polarizations. Notice also that the only non-vanishing term
in ${D_\mu}_{+\times}$ is ${D_0}_{+\times} \equiv D_{+\times}$.

Combining these three pieces together, we finally arrive at
\begin{eqnarray}
I = &{}&\sum_{\vec q}\int a^4 d^4x\Bigg\{
	\frac{1}{2}\sum_s \left(\nabla_\mu \bar h_s \nabla^\mu h_s
	- \frac{m^2_s}{a^2}\bar h_s h_s\right)
\nonumber \\
	&{}&
	- \frac{1}{a^2}D_{\times +}\bar h_{+}\frac{dh_{\times}}{d\eta}
	- \frac{1}{a^2}D_{+\times}\bar h_{\times} \frac{dh_{+}}{d\eta}
	+\frac{N^2_{+\times}}{2}\left(\bar h_{+}h_\times
	+\bar h_{\times}h_{+}\right)
	\Bigg\}\>,
\label{e46}
\end{eqnarray}
where
\begin{equation}
m^2_{s}(\epsilon)=a^2\left(M^2_{++} + T_{++} - N^2_{ss}\right)\>.
\label{e47}
\end{equation}
Once again it is important to note that we have not, as yet, made any
specific choice of polarization vectors.

At this point we can see the differences between the propagation of
GW in a B-I universe versus a RW universe. Ford and Parker has shown
that in a RW universe the two polarizations of the GW decouple from
one another and each, separately, is equivalent to a minimally
coupled massless scalar field. Neither is true for the GW propagating
in a B-I universe, however. Here the GW picks up an effective mass
term due to it being a spin-2 field and is much more sensitive
to the anisotropy than a scalar field is. What is even more
surprising is the apparent coupling which is present
between the two polarization states of the GW.

The physical relevance of this coupling term is somewhat questionable
at this point, however. They could have arisen from an inappropriate
choice of polarization vectors or, since the second
such term comes from the gauge piece of the action, because we made
an inconvenient gauge choice for the GW. For example even in a RW
universe a coupling term between the polarizations would
appear if we had chosen polarization tensors which were not
orthogonal to one another; an ``inappropriate'' choice of
polarization tensors. To begin to separate the truly physical effects
of the anisotropy on the GW from the arbitrariness in defining the
GW, we have to develope a better understanding of the effects of the
anisotropy on the GW. This is best done by considering the behavior
of the polarization vectors instead of the full tensors.

Consider the triad $(\hat q_j, \epsilon^{(1)}_j, \epsilon^{(2)}_j)$ which
form a local orthonormal coordinate system. Because
we are in the B-I universe, these three vectors are not fixed, but
rather changes with $\eta$. Since they are constrained to lie on the
unit sphere, their motion consists of two rotations; one in
the plane perpendicular to $\hat q$ which is spanned by
$\epsilon^{(1)}_j$ and $\epsilon^{(2)}_j$, the other parallel to
$\hat q$. Let us consider the two rotations
separately. Rotations parallel to $\hat q$ occur because the
direction of propagation of the GW, $\hat q_j$, is
always changing with time (see eq.~$(\ref{e25})$). This is because the
medium through which the GW is propagating, the B-I universe, is
anisotropic and always changing with $\eta$. In
particular, notice that in a RW universe, where such coupling between
polarization states is not present, $\hat q_j$ does not change with
time and there are no rotations along this direction. Rotations in
this direction are therefore caused by a physical effect of the
anisotropy on the GW.

Rotations in the $\epsilon^{(1)}$-$\epsilon^{(2)}$ plane, however, are not
physical. They can occur even when $\hat q_j$ does not change
directions as in a RW universe. Remember that the definition of
$\epsilon^{(s)}_j$ is somewhat arbitrary. One still has the freedom
to do a rotation as in eq.~$(\ref{e22})$ of the polarization vectors in this
plane even if this rotation is $\eta$-dependent. Returning to
eq.~$(\ref{e45})$ we see
that $D_{+\times}=2\epsilon^{(1)}_\mu\nabla_0{\epsilon^{(2)}}^\mu=
-2\epsilon^{(2)}_\mu\nabla_0{\epsilon^{(1)}}^\mu$ depends explicitly on the
velocity $\nabla_0 \epsilon^{(s)}_j$ of the polarization vectors {\it
in this plane}. Consequently, the presence of the $D_{+\times}$
coupling term is due solely to the rotation of the polarization
vectors in the plane perpendicular to $\hat q_j$. A $D_{+\times}\ne0$
therefore only means that we have not chosen the ``correct''
polarization vectors: one in which the polarization vectors do not
rotate about $\hat q_j$.

To demonstrate that such a choice of polarization vectors always
exists, we preform the $\eta$-dependent rotation of the polarization
vectors given in eq.~$(\ref{e22})$. Under this rotation, we find that
\begin{equation}
D_{+\times}(\hat\epsilon) = 2\phi'+D_{+\times}(\epsilon)\>,
\label{e48}
\end{equation}
while
\begin{eqnarray}
m^2_{+}(\hat\epsilon) = &{}&
	m^2_{+}(\epsilon) -4(\phi')^2 -
	8\phi'\sum_j\epsilon^{(1)}_j\nabla_0{\epsilon^{(2)}}^j -
\nonumber \\
	&{}&
	a^2\left\{
	N^2_{++}(\epsilon)(\cos^2{2\phi}-1) +
	N^2_{+\times}(\epsilon)\sin{4\phi}  +
	N^2_{\times\times}(\epsilon)\sin^2{2\phi}
	\right\}\>,
\nonumber \\
m^2_{\times}(\hat\epsilon) = &{}&
	m^2_{\times}(\epsilon) -4(\phi')^2 -
	8\phi'\sum_j\epsilon^{(1)}_j\nabla_0{\epsilon^{(2)}}^j -
\nonumber \\
	&{}&
	a^2\left\{
	N^2_{++}(\epsilon)\sin^2{2\phi} -
	N^2_{+\times}(\epsilon)\sin{4\phi}+
	N^2_{\times\times}(\epsilon)(\cos^2{2\phi}-1)
	\right\}\>,
\nonumber \\
N^2_{+\times}(\hat\epsilon)=&{}&\frac{1}{2}\left(
	N^2_{\times\times}(\epsilon)
	-
	N^2_{++}(\epsilon)
	\right)
	\sin4\phi
	+
	N^2_{+\times}(\epsilon)
	\cos4\phi\>,
\label{e49}
\end{eqnarray}
Clearly, one needs only to choose for any given $\epsilon^{(s)}_j${}
\begin{equation}
\phi = -\frac{1}{2}\int^\eta D_{+\times}(\epsilon) d\eta'+ \phi_0\>,
\label{e50}
\end{equation}
and the coupling term proportional
to $D_{+\times}(\hat\epsilon)$ vanishes. $\phi_0$ is an arbitrary,
{\it constant\/} angle and its presence means that we still have a
degree of freedom to choose our polarization vectors. For the
specific choice of polarization vectors given in eq.~$(\ref{e24})$,
\begin{equation}
\phi =
	-\int^\eta \frac{\hat q_1\hat q_2\hat q_3}{\hat q_1^2+\hat
	q_2^2} \left(\frac{a'_1}{a_1}-\frac{a'_2}{a_2}\right)d\eta'
	+\phi_0\>.
\label{e51}
\end{equation}
Notice also that eq.~$(\ref{e51})$ does not have a symmetry under exchange of
$1$-$2$-$3$ which is a reflection of the fact that once a specific choice
of polarization vectors has been made rotational symmetry is broken.
Other choices of $\epsilon^{(s)}_j$ will only result in different forms
for $\phi$, but $D_{+\times}$ will still vanish.

While $D_{+\times}$ vanishes after an appropriate choice of
polarization vectors is made, this still leaves the coupling term
proportional to $N^2_{+\times}$. This term came from the gauge
dependent term of the original action, however, and is different for
different choices of gauge. It is possible that under other, more
clever gauge choices this term will no longer be present. Its
physical relevance is therefore questionable at this point. There are
two ways to address this problem. The first is to redo
this calculation with a different gauge choice and see if the
coupling terms are still present. This we have done in the {\bf
appendix}. This method, however, has the shortcoming in that one can
never be sure that if there is still some other gauge choice for
which the coupling terms vanish identically. The second way is to
develope a physical argument for or against the relevance of the
physical terms. The final conclusion of this argument should
therefore be valid with any gauge choice. This is the approach we
shall now take.

Consider the function
\begin{equation}
f_{ss'}(\eta)\equiv
\sum_j{\epsilon^{(s)}}^j(\eta_1){\epsilon^{(s')}}_j(\eta)\>,
\label{e51a}
\end{equation}
where $\eta_1$ is some {\it fixed\/} time and
${\epsilon^{(s)}}^j(\eta_1)\equiv
g^{jj}(\eta_1){\epsilon^{(s)}}_j(\eta_1)$. Clearly, $f_{ss'}(\eta_1)
= - \delta_{ss'}$ and thus $f_{ss'}$ is a measure of how the
orthonormality of the polarization vector changes with time. Next,
considering $f_{ss'}$ as a function of $\eta$ only, we expand
$\epsilon_j(\eta)$ in a Taylor series in $\eta$ about $\eta_1$ and
find that to first order,
\begin{equation}
f_{ss'}(\eta) = -\delta_{ss'}
		+
		\left(
		\sum_j{\epsilon^{(s)}}^j\frac{d\epsilon^{(s')}_j}{d\eta}
		\right)
		\Bigg\vert_{\eta=\eta_1}(\eta-\eta_1)+\dots\>.
\label{e51b}
\end{equation}
Since $\eta_1$ is arbitrary, we see that the infinitesimal change in
the normality condition is measured by
\begin{equation}
\sum_j {\epsilon^{(1)}}^j\frac{d\epsilon^{(1)}_j}{d\eta}=
	\sum_j\frac{a_j'}{a_j}{\epsilon^{(1)}}^j\epsilon^{(1)}_j
\>,\qquad
\sum_j {\epsilon^{(2)}}^j\frac{d\epsilon^{(2)}_j}{d\eta}=
	\sum_j\frac{a_j'}{a_j}{\epsilon^{(2)}}^j\epsilon^{(2)}_j\>.
\label{e51c}
\end{equation}
The difference of these two terms is just $\sqrt{-a^2N^2_{++}}=\sum
{\varpi^{(+)}}^j_ja_j'/a_j$ while their sum is just the $\sum
\tau^j_ja_j'/a_j$ defined in eq.~$(\ref{a2})$ of the {\bf appendix}. For
the choice of gauge eq.~$(\ref{e18})$ it plays no role. The
infinitesimal change to the orthogonality condition is measured by
the linear combinations
\begin{equation}
\sum_j
\left({\epsilon^{(1)}}^j\frac{d\epsilon^{(2)}_j}{d\eta}
	+
	{\epsilon^{(2)}}^j\frac{d\epsilon^{(1)}_j}{d\eta}
\right)=
	2\sum_j\frac{a_j'}{a_j}{\epsilon^{(1)}}^j\epsilon^{(2)}_j\>,
\label{e51d}
\end{equation}
which is just
$\sqrt{-a^2N^2_{\times\times}}=\sum{\varpi^{(\times)}}^j_ja_j'/a_j$,
and
\begin{equation}
\sum_j
\left({\epsilon^{(1)}}^j\frac{d\epsilon^{(2)}_j}{d\eta}
	-
	{\epsilon^{(2)}}^j\frac{d\epsilon^{(1)}_j}{d\eta}
\right)= D_{+\times}\>.
\label{51e}
\end{equation}
We thus see that each of the terms which may cause a coupling
between the two polarization states, $N^2_{++}$, $N^2_{\times\times}$,
$N^2_{+\times}$, and $D_{+\times}$, have their roots in the
time rate of change of the polarization vectors. Since the
polarization vectors change with time due to the anisotropic rate of
expansion of a B-I universe, we therefore conclude that the presence
of these coupling terms is a natural consequence of the physical
properties of the spacetime. In particular, we see that a coupling
between the two polarization states will be present so long as we are
not able to consistently define a polarization state which is valid
at all times. They will be present in some form no
matter what gauge conditions one chooses and the coupling terms are
physically relevant.

To further illustrate the connection between the rate of change in
the direction of the polarization vectors and the coupling term, let
us look at the conditions under which theses coupling terms will vanish.
For the special choice of polarization vectors given in
eq.~$(\ref{e24})$,
\begin{eqnarray}
D_{+\times} = &{}& -2\frac{\hat q_1\hat q_2\hat q_3}{\hat q_1^2+\hat q_2^2}
	      \left(\frac{a_1'}{a_1}-\frac{a_2'}{a_2}\right)\>,
\nonumber \\
N^2_{+\times}=&{}&
	\frac{2}{a^2}
	\frac{\hat q_1\hat q_2\hat q_3}{\hat q_1^2+\hat q_2^2}
	      \left(\frac{a_1'}{a_1}-\frac{a_2'}{a_2}\right)
\nonumber \\
	&{}&
	\left[-\hat q_1^2\left(\frac{a_2'}{a_2}-\frac{a_3'}{a_3}\right)
	+
	\hat q_2^2\left(\frac{a_3'}{a_3}-\frac{a_1'}{a_1}\right)
	+
	\frac{(\hat q_1^2-\hat q_2^2)\hat q_3^2}{\hat q_1^2+\hat
	q_2^2}\left(\frac{a_1'}{a_1}-\frac{a_2'}{a_2}
	\right)
	\right]\>.
\label{e51'}
\end{eqnarray}
We caution the reader that while $D_{+\times}$ is invariant under
{\it constant\/} rotations of the polarization vectors,
$N^2_{+\times}$ is not and our arguments are valid only within this
choice $\epsilon^{(s)}_j$. We see that all coupling between
the two linear polarizations will vanish if either $a_1 = a_2$ or if
$q_j=0$ along some direction $j$. Although the first condition seems
to break the $1$-$2$-$3$ exchange symmetry in the labeling of the
anisotropy axis, this simply means that we were fortunate enough to
choose the ``canonical'' polarization vectors for the GW which
decouples the two polarizations. Any other choice of polarization
vectors will produce an $N^2_{+\times}\ne0$, but can always be made
to vanish with a constant rotation of the polarization vectors.
$D_{+\times}$, on the other hand, vanishes for all such choices of
polarization vectors.

Referring to eq.~$(\ref{e24})$ we see that when $a_1=a_2$,
$\epsilon^{(2)}_j$ lies in the $1$-$2$ plane and, more importantly,
its direction does not change with time; $\nabla_0\epsilon^{(2)}_j$
is once again parallel to $\epsilon^{(2)}$. Thus any direction we
choose for this polarization will always lie in this direction and
will not change with time.

Suppose, now that $a_1\ne a_2$, but $q_3=0$. Then all the coupling
terms between the two polarization states vanish. In this case $\hat
q_j$ lies in the $1$-$2$ plane but more importantly we may always choose
one of the polarization vectors to lie in the $3$-direction
perpendicular to this plane. Once again we see that the direction of
this polarization vector does not change and we can once again have a
consistant definition of polarization vectors for all time.

We now see explicitly that the coupling terms vanish as long as the
direction of one of the polarization vectors does not change with
time. This also can be seen in the definition of $N^2_{+\times}$
which depends on the directional derivative of the polarization
vectors. Physically, it means that when this happens we can
consistantly define the direction of at least one of the polarization
vectors at all times. The second polarization can be found by taking
the cross product of this polarization vector with the direction of
propagation of the GW. Although this can be done in certain special
cases, it is not true in general and there will in general be a
coupling between polarization states of the GW no matter which gauge
one picks.

Finally, we note that for circularly polarized GW
\begin{equation}
I = \sum_{\vec q}\int a^4d^4x\left\{
	\sum_{s=R,L} \frac{1}{2}
	\left(
		\nabla_\rho \bar h_s\nabla_\rho h_s
		-
		\frac{m^2(\hat\epsilon)}{a^2}\bar h_s h_s
	\right)
	+
	\frac{\beta^2(\hat\epsilon)}{a^2}\bar h_R h_L
	+
	\frac{\bar\beta^2(\hat\epsilon)}{a^2}h_R \bar h_L
	\right\}
\label{e52}
\end{equation}
where $h_R = (h_{+}-ih_\times)/\sqrt 2$, $h_L =
(h_{+}+ih_\times)/\sqrt 2$,
\begin{equation}
\beta(\hat\epsilon) = e^{2i\phi}\beta(\epsilon)\>,
\label{e53}
\end{equation}
with
\begin{equation}
\beta(\epsilon) = \frac{1}{2}\left[
		  \epsilon^{(1)}_\mu\nabla^\mu\epsilon^{(2)}_0
		  +
		  \epsilon^{(2)}_\mu\nabla^\mu\epsilon^{(1)}_0
		  +i(\epsilon^{(1)}_\mu\nabla^\mu\epsilon^{(1)}_0
		  -
		  \epsilon^{(2)}_\mu\nabla^\mu\epsilon^{(2)}_0
		  )
		  \right]\>,
\label{e54}
\end{equation}
and $\phi$ defined through eq.~$(\ref{e50})$. We have thus
eliminated the $D_{+\times}$ with this choice of polarization
vectors. Notice that for circularly polarized GW the mass terms for
both polarizations are equal
\begin{eqnarray}
m^2(\hat\epsilon) &=&
\frac{1}{2}[m^2_{+}(\hat\epsilon)+m^2_{-}(\hat\epsilon)]
\nonumber \\
    &=&
    -\sum_{cyclic}
    \left(
    	\frac{a_j'}{a_j} - \frac{a_k'}{a_k}
    \right)^2\hat q^2_l
    -\sum_{j>k}
    \left(
    	\frac{a_j'}{a_j} - \frac{a_k'}{a_k}
    \right)^2\hat q^2_j \hat q^2_k
    +\frac{a^2}{2}(N^2_{++}(\hat\epsilon)+N^2_{\times\times}(\hat\epsilon))\>,
\label{e55}
\end{eqnarray}
but we pay for this symmetry through the addition of a complex
coupling term between the two polarization states. This coupling term
will only vanish when both the real and imaginary parts of
$\beta(\epsilon)$ vanish. If, however, they vanish for one choice
of polarization vectors, they will vanish for any other choice from
eq.~$(\ref{e53})$. Consequently, we may use the polarization vectors
given in eq.~$(\ref{e25})$, and find that
\begin{equation}
0 = \epsilon^{(1)}_\mu\nabla^\mu\epsilon^{(2)}_0 +
\epsilon^{(2)}_\mu\nabla^\mu\epsilon^{(1)}_0 =
\frac{\hat q_1\hat q_2\hat q_3}{\sqrt{\hat q_1^2 +\hat q_2^2}}
\left(\frac{a'_1}{a_1}-\frac{a'_2}{a_2}\right)\>,
\label{e56}
\end{equation}
as well as
\begin{equation}
0= \epsilon^{(1)}_\mu\nabla^\mu\epsilon^{(1)}_0 -
\epsilon^{(2)}_\mu\nabla^\mu\epsilon^{(2)}_0 =
-\hat q_1^2\left(\frac{a_2'}{a_2}-\frac{a_3'}{a_3}\right)
+\hat q_2^2\left(\frac{a_3'}{a_3}-\frac{a_1'}{a_1}\right)
+
\frac{(\hat q_1^2-\hat q_2^2)\hat q_3^2}{\hat q_1^2+\hat
q_2^2}\left(\frac{a_1'}{a_1}-\frac{a_2'}{a_2}\right)\>,
\label{e57}
\end{equation}
must both vanish separately. This can only happen when the GW
propagates along an asymmetry axis and the spacetime is axially
symmetric in the plane perpendicular to this axis. For example,
the coupling term will vanish if $\hat q_3 =-1$ and $a_1=a_2$. This
is a much more stringent condition than for linearly polarized GW and
is because although the two linearly polarized GW may decouple from one
another, they will still have different mass terms and this
introduces an additional coupling. Thus we have the peculiar
situation that while the linear polarized states may decoupled
from one another, the circularly polarized states need not. This
once again underscores the fact that in a general B-I universe one cannot
consistently define a polarization state at all times.

\noindent{\bf \S 5. Solutions of Equation of Motion}

The evolution equation for GW in an arbitrary B-I spacetime cannot be
solved in terms of known functions in general, especially when there
is a coupling between the two polarization states. Consequently, we
shall only consider GW propagating in either the Kasner or the
Zel'dovich spacetimes where solutions {\it can\/} be found in certain
special cases.

{}From the action eq.~$(\ref{e46})$ the evolution equation for a
linearly polarized GW propagating in a Kasner or Zel'dovich spacetime is
\begin{eqnarray}
0&=&
   \frac{d^2h_{\vec q +}}{dy^2}+2\frac{1}{a}\frac{da}{dy}
   \frac{dh_{\vec q +}}{dy} +
   \left\{
   	\sum_{j=1}^3y^{1-3p_j}\widetilde q^2_j +\widetilde m_{+}^2
   \right\}h_{\vec q +}
   +
   2\widetilde D_{+\times}\frac{dh_{\vec q\times}}{dy}
   -a^2\widetilde N^2_{+\times}h_{\vec q\times} \>,
\nonumber \\
0&=&
   \frac{d^2h_{\vec q\times}}{dy^2}+2\frac{1}{a}\frac{da}{dy}
   \frac{dh_{\vec q\times}}{dy}+
   \left\{
       \sum_{j=1}^3y^{1-3p_j}\widetilde q^2_j +\widetilde m_{\times}^2
   \right\}h_{\vec q\times}
   +
   2\widetilde D_{+\times}\frac{dh_{\vec q +}}{dy}
   -a^2\widetilde N^2_{+\times}h_{\vec q +} \>,
\nonumber \\
&{}&
\label{e58}
\end{eqnarray}
where we have not chosen a $D_{+\times}=0$ and we have taken $y_0=1$
for convenience. In the above, the tilde
denotes the fact that we have scaled the corresponding quantity by
either $4Q_0$ or $4(8\pi\rho_0/3+Q_0)$
depending on whether the spacetime is a Kasner or a Zel'dovich
universe to make the resulting quantity
dimensionless. For example, $\widetilde q_j^2 \equiv q_j^2/4/Q_0$ for the
Kasner universe while $\widetilde q_j^2 \equiv
q_j^2/4/(8\pi\rho_0/3+Q_0)$ for the Zel'dovich universe.
We see that due to the coupling term between the two polarizations,
even if the GW is initially plus-polarized, a cross polarization will
be generated. Unfortunately, even in the case of the Kasner universe
eq.~$(\ref{e58})$ cannot be solved generally. We shall instead have
to look at limiting conditions.

Suppose, for convenience, that $p_1>p_2>p_3$. Let us now
consider a general $\vec q$ for which each of the components of $\vec
q$ do not vanish. If we then take the limit $y\to0$ we find that $\hat
q_1 = -1$ while $\hat q_2 = \hat q_3=0$. With the choice of
polarization vectors given in eq.~$(\ref{e22})$,
\begin{equation}
\widetilde m^2_{+} = 0\qquad,\quad \widetilde m^2_\times = -
	\left(\frac{a_2'}{a_2}-\frac{a_3'}{a_3}\right)^2\>,
\label{e59}
\end{equation}
where the prime now denotes the derivative with respect to $y$.
$N_{+\times}=D_{+\times}=0$ and there is no coupling between
the two polarizations. Similarly, when $y\to\infty$, $\hat q_1=\hat
q_2=0$ while $\hat q_3 =-1$. Now
\begin{equation}
\widetilde m^2_{+} = 0\qquad,\quad \widetilde m^2_\times = -
	\left(\frac{a_1'}{a_1}-\frac{a_2'}{a_2}\right)^2\>.
\label{e60}
\end{equation}
In these two limits the behavior of the GW
simplifies dramatically and can be solved in closed form. More
importantly, these two limits, which corresponds to the behavior of
the GW near the initial singularity as well as the large time behavior
of the GW, are the most physically significant.

It is therefore sufficient to consider a GW propagating along one of the
symmetry axis; say in the $l$th direction. Then
in eq.~$(\ref{e58})${} $h_{\vec q+}$ and $h_{\vec q\times}$ decouple
into two equations
\begin{eqnarray}
0&=&
   \frac{d^2h_{\vec q +}}{dy^2}+2\frac{1}{a}\frac{da}{dy}
   \frac{dh_{\vec q +}}{dy}+y^{1-3p_l}\widetilde q^2_lh_{\vec q +}\>,
\nonumber \\
0&=&
   \frac{d^2h_{\vec q\times}}{dy^2}+2\frac{1}{a}\frac{da}{dy}
   \frac{dh_{\vec q\times}}{dy}+
   \left\{
      y^{1-3p_l}\widetilde q^2_l -
      \left(
      	\frac{a_j'}{a_j}-\frac{a_k'}{a_k}
      \right)^2
   \right\}h_{\vec q\times}\>,
\label{e61}
\end{eqnarray}
where $l\ne j\ne k$. Notice that in general the behavior of $h_{\vec q +}$
will be different from that of $h_{\vec q\times}$ due to the additional
masslike term in its equation of motion. It is only when we are
propagating along the asymmetry axis of an axially symmetric B-I
universe will the two polarizations have the same behavior. Since the
solution for $h_{\vec q +}$ can be obtain from that of $h_{\vec
q\times}$ by setting $a_j=a_k$ in the mass term, we shall concentrate
our attention on the solution of $h_{\vec q\times}$.

It is straightforward to see that the solution to eq.~$(\ref{e61})$
are Bessel functions $J_\nu$ and $Y_\nu$. Using the parametrization of
the $p_j $ given in eq.~$(\ref{e12})$, we find that
\begin{eqnarray}
h_{\times}^{(1)}(\vec q,x) = e^{iq_1 x^1}
	\Bigg\{
	&A_\times^{(1)}&
	J_{\nu_{+}}\left(
		\frac{\widetilde q_1 y^{1-\frac{1}{2}\cos\theta-
		\frac{\sqrt3}{2}\sin\theta}}
		{1-\frac{1}{2}\cos\theta-\frac{\sqrt3}{2}\sin\theta}
	\right)
	+
\nonumber \\
	&B_\times^{(1)}&Y_{\nu_{+}}\left(
		\frac{\widetilde q_1 y^{1-\frac{1}{2}\cos\theta-
		\frac{\sqrt3}{2}\sin\theta}}
		{1-\frac{1}{2}\cos\theta-\frac{\sqrt3}{2}\sin\theta}
	\right)
	\Bigg\}\>,
\nonumber \\
h_{\times}^{(2)}(\vec q,x) = e^{iq_2 x^2}
	\Bigg\{
	&A_\times^{(2)}&J_{\nu_{-}}\left(\frac{\widetilde q_2
	y^{1-\frac{1}{2}\cos\theta+\frac{\sqrt3}{2}\sin\theta}}
	{1-\frac{1}{2}\cos\theta+\frac{\sqrt3}{2}\sin\theta}\right)
	+
\nonumber \\
	&B_\times^{(2)}&Y_{\nu_{-}}\left(\frac{\widetilde q_2
	y^{1-\frac{1}{2}\cos\theta+\frac{\sqrt3}{2}\sin\theta}}
	{1-\frac{1}{2}\cos\theta+\frac{\sqrt3}{2}\sin\theta}
	\right)
	\Bigg\}\>,
\nonumber \\
h_{\times}^{(3)}(\vec q,x) = e^{iq_3 x^3}
	\Bigg\{
	&A_\times^{(3)}&J_{\nu_3}\left(\frac{\widetilde q_3
	y^{1+\cos\theta}}
	{1+\cos\theta}\right)
	+
\nonumber \\
	&B_\times^{(3)}&Y_{\nu_{3}}\left(\frac{\widetilde q_3
	y^{1+\cos\theta}}
	{1+\cos\theta}
	\right)
	\Bigg\}\>,
\label{e62}
\end{eqnarray}
where
\begin{eqnarray}
\nu_{\pm} &\equiv& \frac{\vert3 \cos\theta\mp\sqrt{3}\sin\theta\vert}
		{2-\cos\theta\mp\sqrt3\sin\theta}\>,
\nonumber \\
\nu_3
&\equiv&
\sqrt{3}\frac{\vert\sin\theta\vert}{1+\cos\theta}\>,
\label{e63}
\end{eqnarray}
while $A_\times^{(j)}$ and $B_\times^{(j)}$ are integration constants. The
superscript $h^{(j)}_{\vec q\times}$ signifies that this is a GW
which propagates along the $j$th direction. The $h^{(j)}_{\vec q +}$
solutions are obtained from the above by replacing the Bessel
functions of various orders with Bessel functions of order zero but
with the same arguments.

Of particular interest is the small and large $y$ limits to
eq.~$(\ref{e62})$. The large $y$ limit determines
whether or not the GW (which is a perturbation of the metric)
increases or decreases in magnitude. It thereby establishes
whether or not the Kasner and Zel'dovich universe are stable under
small perturbations. The small $y$ limit, on the other hand,
determines the behavior of the GW near the initial singularity. To do
so, however, we shall have to find a way of comparing matrices. We
therefore define the norm
\begin{equation}
\Vert T_{\mu\nu}\Vert \equiv \hbox{max}\>\vert T_{\mu\nu}\vert\>,
\label{e63'}
\end{equation}
where the max is taken over all components of the tensor. This norm
has the advantage of having all the nice properties associated with a
norm $\cite{com2}$.

Taking first the $y\to\infty$ limit, we find that
\begin{eqnarray}
h_{\times}^{(1)}(\vec q,x) &\approx& e^{iq_1x^1}
		\sqrt{\frac{2-\cos\theta-\sqrt3\sin\theta}
		{\widetilde q_1\pi}}
		y^{-(2-\cos\theta-\sqrt3\sin\theta)/4}
\nonumber \\
	&{}&
		\Bigg\{
		A_\times^{(1)}
		\cos\Bigg(
		\frac{2\widetilde q_1 y^{(2-\cos\theta-\sqrt3\sin\theta)/2}}
		{2-\cos\theta-\sqrt3\sin\theta}
\nonumber \\
	&{}&
		-\frac{\pi}{2}\left(\nu_{+} +\frac{1}{2}\right)
		\Bigg)
		+
		B_\times^{(1)}
		\sin\Bigg(
		\frac{2\widetilde q_1y^{(2-\cos\theta-\sqrt3\sin\theta)/2}}
		{2-\cos\theta-\sqrt3\sin\theta}
		-\frac{\pi}{2}\left(\nu_{+}+\frac{1}{2}\right)
		\Bigg)
		\Bigg\}\>,
\nonumber \\
	&{}&
\end{eqnarray}
while
\begin{eqnarray}
h_{\times}^{(3)}(\vec q,x) &\approx& e^{iq_3x^3}\sqrt{\frac{2(1+\cos\theta)}
		{\widetilde q_1\pi}}y^{-(1+\cos\theta)/2}
		\Bigg\{
		A_\times^{(3)}\cos\left(\frac{
		\widetilde q_3y^{1+\cos\theta}
		}{1+\cos\theta}
		-\frac{\pi}{2}\left(\nu_{3}+\frac{1}{2}\right)\right)
		+
\nonumber \\
		&{}&
		B_\times^{(3)}\sin\left(\frac{
		2\widetilde q_3y^{1+\cos\theta}
		}{1+\cos\theta}
		-\frac{\pi}{2}\left(\nu_{3}+\frac{1}{2}\right)\right)
		\Bigg\}\>,
\end{eqnarray}
with $h_{\vec q\times}^{(2)}$ being obtained from $h_{\vec
q\times}^{(1)}$ by taking $\theta\to-\theta$ and replacing $(1)$ with
$(2)$ everywhere. It is then straightforward to show that the
amplitude of $\Vert h^{(s)}_{\mu\nu}\Vert /\Vert g_{\mu\nu}\Vert$
decreases with increasing $y$. Consequently, we see that the
perturbation $h_{\mu\nu}$ of the metric decreases with increasing
$y$.

Taking now the small $y$ limit, we find that
\begin{eqnarray}
h_{\vec q\times}^{(1)} &\sim&
y^{-\frac{3}{2}\vert\cos\theta-\frac{\sqrt3}{3}\sin\theta\vert}\>,
\nonumber \\
h_{\vec q\times}^{(3)} &\sim& y^{-\sqrt 3\sin\theta}\>,
\label{e64}
\end{eqnarray}
where once again $h_{\vec q\times}^{(2)}$ is obtained from $h_{\vec
q\times}^{(1)}$ by taking
$\theta\to-\theta$. Eq.~$(\ref{e64})$ holds as long as the degree of
the divergence in $h_{\vec q\times}^{(j)}$ does not vanish. When it
does, that $h_{\vec q\times}^{(j)}$ will have a logarithmic
divergence. We therefore find that as long as $p_2\ne p_3$,
$\Vert h^{(\times)}_{\mu\nu}\Vert/\Vert g_{\mu\nu}\Vert \sim 1$ from
GW propagating along the $1$ or $3$ direction while $\Vert
h^{(\times)}_{\mu\nu}\Vert/\Vert g_{\mu\nu}\Vert \sim 0$ for a GW
propagating along the $2$ direction as $y\to0$. Similarly, $\Vert
h^{(+)}_{\mu\nu}\Vert/\Vert g_{\mu\nu}\Vert \sim \log y$ for a GW
propagating along the $1$ or $2$ directions and $\Vert
h^{(+)}_{\mu\nu}\Vert/\Vert g_{\mu\nu}\Vert\to0$ for a  GW
propagating along the $3$ direction. We thus see that along
the $1$ and $2$ directions the perturbation of the metric becomes
unboundly large near the singularity.

For the special case of $p_2=p_3$, we find that $\Vert
h^{(\times)}_{\mu\nu}\Vert/\Vert g_{\mu\nu}\Vert \sim 1$ for GW's
propagating along the $2$ or $3$ directions while $\Vert
h^{(\times)}_{\mu\nu}\Vert/\Vert g_{\mu\nu}\Vert \sim \log y$ for a
GW propagating along the $1$ direction. Similarly, we find that $\Vert
h^{(+)}_{\mu\nu}\Vert/\Vert g_{\mu\nu}\Vert \sim \log y$ for a GW
propagating along the $1$ or $2$ direction, while $\Vert
h^{(\times)}_{\mu\nu}\Vert/\Vert g_{\mu\nu}\Vert \to0$ for a GW
propagating along the $3$ direction. Thus, in this special case the
perturbation of the metric becomes unboundly large along all three
directions. The case considered by Hu, which corresponds to
$\theta=0$, belongs to this catagory and the results we have obtained
agree with his.

The case where $\theta =\pi/3$ is quite special since this
corresponds to $p_2=p_3=0$ while $p_1=1$. Supposedly, this case also
falls within the above analysis and we would still obtains the same
small $y$ behavior. When $\rho_0=0$ this is the Rindler spacetime,
however, which is known to be equivalent to Minkowski space and
supposedly the propagation of GW in a Minkowski spacetime does not
have a singularity. To address this problem, we now perform a more
detailed analysis of the solution in this special case.

Going back to the equation of motion, we see that when
$\theta=\pi/3$,
\begin{eqnarray}
h^{(+)}_{\mu\nu}(q_1,x) &\equiv&
		\varpi^{(+)}_{\mu\nu}(q_1)h^{(1)}_{+}
\nonumber \\
		&=&
		\left(e^3_\mu e^3_\nu-e^2_\nu e^2_\mu\right)
		\left(
			A^{(1)}_{+}e^{iq_1(x^1+t_0\log t/t_0)}
			+
			B^{(1)}_{+}e^{iq_1(x^1-t_0\log t/t_0)}
		\right)\>,
\nonumber \\
h^{(+)}_{\mu\nu}(q_2,x) &\equiv&
		\varpi^{(+)}_{\mu\nu}(q_2)h^{(2)}_{+}
\nonumber \\
		&=&
		\left(e^3_\mu e^3_\nu-e^1_\nu e^1_\mu\right)
		\left[
			A^{(2)}_{+} J_0\left(q_2t\right)
			+
			B^{(2)}_{+} Y_0\left(q_2t\right)
		\right]e^{iq_2x^2}\>,
\nonumber \\
h^{(+)}_{\mu\nu}(q_3,x) &\equiv&
		\varpi^{(+)}_{\mu\nu}(q_3)h^{(3)}_{+}
\nonumber \\
		&=&
		\left(e^2_\mu e^2_\nu-e^1_\nu e^1_\mu\right)
		\left[
			A^{(3)}_{+} J_0\left(q_3t\right)
			+
			B^{(3)}_{+} Y_0\left(q_3t\right)
		\right]e^{iq_3x^3}\>,
\label{e65}
\end{eqnarray}
where we have put back in the polarization tensors explicitly, $t_0 =
2\eta_0/3$ and we have used $y= (t/t_0)^{2/3}$ for the Rindler
spacetime. Similarly,
\begin{eqnarray}
h^{(\times)}_{\mu\nu}(q_1,x) &\equiv&
		\varpi^{(\times)}_{\mu\nu}(q_1)h^{(1)}_{\times}
\nonumber \\
		&=&
		\left(e^2_\mu e^3_\nu+e^2_\nu e^3_\mu\right)
		\left(
			A^{(1)}_{\times}e^{iq_1(x^1+t_0\log t/t_0)}
			+
			B^{(1)}_{\times}e^{iq_1(x^1-t_0\log t/t_0)}
		\right)\>,
\nonumber \\
h^{(\times)}_{\mu\nu}(q_2,x) &\equiv&
		\varpi^{(\times)}_{\mu\nu}(q_2)h^{(2)}_{\times}
\nonumber \\
		&=&
		-\left(e^1_\mu e^3_\nu+e^1_\nu e^3_\mu\right)
		\left[
			A^{(2)}_{\times} J_1\left(q_2t\right)
			+
			B^{(2)}_{\times} Y_1\left(q_2t\right)
		\right]e^{iq_2x^2}\>,
\nonumber \\
h^{(\times)}_{\mu\nu}(q_3,x) &\equiv&
		\varpi^{(\times)}_{\mu\nu}(q_3)h^{(3)}_{\times}
\nonumber \\
		&=&
		-\left(e^1_\mu e^2_\nu+e^1_\nu e^2_\mu\right)
		\left[
			A^{(3)}_{\times} J_1\left(q_3t\right)
			+
			B^{(3)}_{\times} Y_1\left(q_3t\right)
		\right]e^{iq_3x^3}\>,
\label{e66}
\end{eqnarray}
and we can see explicitly the singularity in the solutions when
$t\to0$.

Let us now do a coordinate transformation into Minkowski spacetime.
When $\theta=\pi/3$, the metric is
\begin{equation}
ds^2 = (dt)^2-\left(\frac{t}{t_0}\right)^2(dx^1)^2 -(dx^2)^2-(dx^3)^2\>.
\label{e67}
\end{equation}
To map this into the Minkowski spacetime, we make the coordinate
transformation
\begin{equation}
T=t\cosh(x^1/t_0)\>,\quad X^1=t\sinh(x^1/t_0)\>,\quad X^2=x^2\>,
\quad X^3=x^3\>.
\label{e68}
\end{equation}
Notice, however, that because $h_{\mu\nu}$ is a tensor, it transforms
as
\begin{equation}
_Mh_{\mu\nu} = \frac{\partial x^\alpha}{\partial X^\mu}
		\frac{\partial x^\beta}{\partial
		X^\nu}h_{\alpha\beta}\>,
\label{e69}
\end{equation}
where $_Mh_{\mu\nu}$ is the transformed GW. Note, however, that
$\partial x^\alpha/\partial X^\mu$ is block diagonal and mixes the
$0$-$1$ components of the GW while leaving the $2$-$3$ components
alone. For GW propagation in the $2$ or $3$ directions, this means
that the transformed $_Mh_{0\nu}\ne0$, although $q^j _Mh_{ij}=0$
still and it will still be traceless. Consequently, for GW
propagating in these two directions, the transformed GW will not
satisfy the usual synchronous gauge and is not what one usually calls
a GW in a Minkowski spacetime.

For GW propagating along the $1$-axis, however, the situation is a
little different. Notice first that the polarization matrices will
not be changed by the transformation and that they are, in fact,
precisely the polarization matrices for a GW propagating in this
direction in a Minkowski spacetime. Next, using the coordinate
transeformation eq.~$(\ref{e68})$, we see that
\begin{equation}
h^{+}_{\mu\nu}(q_1, x) = \varpi_{\mu\nu}^{+}(q_1)
			\left\{
			A^{(1)}_{+}
			\left(\frac{T+X^1}{t_0}\right)^{iq_1t_0}
			+
			B^{(1)}_{+}
			\left(\frac{T-X^1}{t_0}\right)^{-iq_1t_0}
			\right\}\>,
\label{e70}
\end{equation}
and the equation for the $\times$ mode has the same form. Notice that
eq.~$(\ref{e70})$ is a solution of the wave equation in Minkowski
spacetime, as a GW in a Minkowski spacetime should. In
this sense it is a GW in the Minkowski spacetime, but it is
not plane wave and satisfies different boundary conditions. Consequently,
we see that after
a coordinate transformation into a Minkowski spacetime, the GW in the
Rindler spacetime will not be transformed into what we usually refer
to as a GW in the Minkowski spacetime.

\noindent{\bf \S 6. Concluding Remarks}

To conclude, we have begun the analysis of the propagation of GW in
B-I universes using the method developed by Ford and Parker. We find
the behavior of a GW in a B-I universe to be very much
different than a GW in a RW universe. The
two polarization states are {\it not\/} equivalent to two minimally
coupled, massless scalar fields. Rather, {\it each\/} polarization
state gains what is effectively a time dependent mass term
due to the tensorial nature of the GW. Namely, the GW is a spin-2
particle instead of a scalar particle and is consequently much more
sensitive to any anisotropy in the universe. We have also found
the two polarization states to be coupled to one another. This
coupling term depends explicitly on the gauge one picks for the GW and
varies as one varies the choice of gauge. No matter
what gauge choice one makes, however, a coupling between the two
polarization states will always be present and the coupling is not a
gauge artifact.

The reason for this is fairly straightforward. We can consider the
propagation of a GW in a B-I universe as the propagation
of a wave through an anisotropic medium. Although the polarization
and direction of propagation of this GW is initially arbitrary, as it
propagates through the anisotropic medium the medium itself will tend
to change the direction of propagation, and thus the direction of
polarization, of the wave. This can be seen explicitly in a Kasner
universe where the anisotropic medium gradually forces the wave to
propagate only along one of the asymmetry axis. The
anisotropy of the medium is constantly changing with time,
however, and thus the polarization state that the medium wishes the
GW to adopt also constantly changes. This is done through the
coupling term between the polarization states and is caused by
changing of the polarization vectors with time.

We would therefore expect a coupling between the two modes of the GW
to be present in any anisotropic expanding universe. Indeed, such a
coupling has also been found by Ezawa and Soda $\cite{ES}$ who
analysed the effects of the topology on the propagation of GW's. They
considered GW's propagating on plain symmetric spacetimes with two of
the spacetime directions compactified into a torus and also found a
coupling between the plus and cross modes of the GW.

Hu considered a B-I spacetime for which $p_1= p_2= 2/3$ while
$p_3=-1/3$ (we have changed slightly the notation used in
$\cite{Hu}$). Moreover, he only considered a GW propagating along
the cylindrical axis. We have seen, however, that the behavior of GW's
in this spacetime is quite special and atypical. In fact, it is only
for a GW propagating in this manner in this spacetime that the two
linear as well as circular polarizations decouple from one another.
Propagation along any other direction in the spacetime will
introduce a coupling between the two polarization states.

Miedema and van Leeuwen have also analysed the propagation of GW in
B-I universes within a general analysis of perturbations of a B-I
universe. Looking at the GW component of the perturbations, they
found that while a GW can be defined as being transversal at any
given time, as the wave evolves in
time non-transversal components of the wave will begin to appear.
They thereby conclude that GW in B-I universes are in general
non-transversal. The longitudinal components of the
wave, however, were considered to have no physical meaning since they
can be gauged away at any given time. We did not encounter any such
subtleties in our analysis. We have instead found that {\it with the
correct gauge choice\/} the GW will always be transverse.

The difference between our two results can be found in the gauge
conditions that we have taken. Miedema and van Leeuwen has chosen to
work in the synchronous and traceless gauges (see eqs.~$(38)$ and
$(108)$ of $\cite{Mie}$). They then attempted to enforce the usual
transverse condition on the GW (eq.~$(114)$ of $\cite{Mie}$). As we
have shown, these three gauge conditions are incompatible and one of
them must be modified. By imposing the synchronous and traceless
conditions on the evolution equations for the GW (eq.~$(135)$ of
$\cite{Mie}$), they cannot then choose the usual transverse
condition. This inconsistancy may be the root cause for the generation
of the longitudinal components as the wave propagates that they
observed.

\begin{center}
{\bf Acknowledgements}
\end{center}

HTC was supported by the R.O.C. NSC Grant No.
NSC84-2112-M-032-003 while ADS was support by the R.O.C. NSC
Grant No. NSC84-2112-M-001-022.
\pagebreak
\begin{appendix}{}

In this appendix we shall repeat the analysis found in $\bf\S 4$ using
the usual transverse-synchronous gauge in order to show that coupling
terms between the polarization states are still present in this
gauge. Indeed, we shall see that the problem gets worse, not better,
with this gauge choice.

Let us denote the polarization tensor for the GW in this gauge by
$\widehat{\varpi}^{(s)}_{jk}$. Then, so as to keep a basis with which
we can compare $\widehat{\varpi}^{(s)}_{jk}$ with $\varpi^{(s)}_{jk}$,
we shall not change the definition of the polarization vectors.
Rather, we shall once again express $\widehat{\varpi}^{(s)}_{jk}$ in
terms of $\epsilon^{(s)}_j$ by taking
\begin{equation}
\widehat{\varpi}^{(s)}_{jk} \equiv \alpha^{(s)} \varpi^{(+)}_{jk}
		+ \beta^{(s)} \tau_{jk}+
		\gamma^{(s)}\varpi^{(\times)}_{jk} \>,
\label{a1}
\end{equation}
where
\begin{equation}
\tau_{jk} = \epsilon^{(1)}_j\epsilon^{(1)}_k + \epsilon^{(2)}_j
\epsilon^{(2)}_k\>,
\label{a2}
\end{equation}
and $\alpha^{(s)}$, $\beta^{(s)}$, $\gamma^{(s)}$ are coefficients
which are to be determined. Clearly $\widehat{\varpi}^{(s)}_{jk}$
still satisfies the spatial components of the transverse condition.
We still require that the two polarization tensors for the two
polarization states of the GW be orthonormal to one another and this
gives the constraint,
\begin{equation}
1 = (\alpha^{(s)})^2 + (\beta^{(s)})^2 + (\gamma^{(s)})^2\>,
\label{a3}
\end{equation}
The $0$-component of the transversality condition in
eq.~$(\ref{e17})$ can now be used with eq.~$(\ref{a3})$ to determine
$\widehat{\varpi}^{(s)}_{jk}$ uniquely,
\begin{eqnarray}
\widehat{\varpi}^{(+)}_{jk} &=& \cos\varphi \,\varpi^{(+)}_{jk}
			-\sin\varphi \,\tau_{jk}\>,
\nonumber \\
\widehat{\varpi}^{(\times)}_{jk} &=& -\sin\varphi \cos\vartheta\,
			\varpi^{(+)}_{jk}-\cos\varphi\cos\vartheta\,\tau_{jk}
			+\sin\vartheta \,\varpi^{(\times)}_{jk}\>,
\label{a4}
\end{eqnarray}
where
\begin{eqnarray}
\sum_j\frac{a_j'}{a_j} \tau_j^j &\equiv&
\varrho\cos\varphi\sin\vartheta \>,
\nonumber \\
\sum_j\frac{a_j'}{a_j} {\varpi^{(+)}}^j_j &\equiv&
\varrho\sin\varphi\sin\vartheta \>,
\nonumber \\
\sum_j\frac{a_j'}{a_j} {\varpi^{(\times)}}^j_j &\equiv&
\varrho\cos\vartheta \>,
\label{a5}
\end{eqnarray}
and
\begin{equation}
\varrho^2=
\left(\sum_j\frac{a_j'}{a_j} {\varpi^{(+)}}^j_j\right)^2
+
\left(\sum_j\frac{a_j'}{a_j} \tau_j^j \right)^2
+
\left(\sum_j\frac{a_j'}{a_j} {\varpi^{(\times)}}^j_j \right)^2
\label{a6}
\end{equation}
For a RW universe, $\vartheta=\pi/2$ while $\varphi =0$.

At this point, we make a few observations. First, the angles
$\vartheta$ and $\varphi$ are determined precisely by linear
combinations of the terms found in eqs.~$(\ref{e51c})$ and
$(\ref{e51d})$. Second, these terms, which measures the infinitesimal
change in the orthonormality conditions for the polarization vectors,
did not disappear when we change to this gauge choice. This
underscores the fact that they are physically relevant. Third,
the trace of the polarization tensors no longer vanish,
\begin{equation}
\sum_j {\widehat{\varpi}^{(+)}}{}^j_j = 2\sin\varphi\>,\qquad
\sum_j {\widehat{\varpi}^{(\times)}}{}^j_j =
2\cos\varphi\cos\vartheta\>,
\label{a7}
\end{equation}
which will introduce additional terms to the GW lagrangian.

In this gauge, the lagrangian for the GW now becomes
\begin{eqnarray}
\widehat I = \frac{1}{4}\int \sqrt{-g}d^4x
    \Bigg(&{}&
    	\nabla_\mu h_\alpha^\beta\nabla^\mu h^\alpha_\beta +
	8\pi(\rho-p)h_\alpha^\beta h^\alpha_\beta
	- 2R_\beta^\mu h^\alpha\mu h_\alpha^\beta
\nonumber \\
	&{}&
	-2{R^{\mu\beta}}_{\alpha\nu} h_\mu^\nu h^\alpha_\beta
	-\nabla_\rho h\nabla^\rho h -  h R^{\mu\nu}h_{\mu\nu}
	\Bigg)\>,
\label{a8}
\end{eqnarray}
where $h = h_\mu^\mu$ is the trace of $h_{\mu\nu}$ which does not
vanish in this gauge and we are denoting the different choice of
gauge with a hat. Once again we can divide this action up into three
terms, the only difference being that the form of the gauge dependent
piece is now different,
\begin{equation}
\widehat I_g \equiv \frac{1}{4}\int \sqrt{-g}d^4x \left(-\nabla_\rho
h\nabla^\rho h - h R^{\mu\nu}h_{\mu\nu}\right)\>.
\label{a9}
\end{equation}
Proceeding in exactly the same way as before, we find that
\begin{equation}
\widehat I_{K} = \frac{1}{2}\sum_{\vec q}\int a^4 d^4x \left(
		\sum_s \nabla_\rho\bar h_s\nabla^\rho h_s
		-
		\sum_{ss'}\widehat M_{ss'}h_s\bar h_{s'}
		-2\sum_{ss'}\widehat D_{\rho ss'} \bar
		h_{s'}\nabla^\rho h_s
		\right)\>,
\label{a10}
\end{equation}
where $\widehat M_{ss'}$ and $\widehat D_{\rho ss'}$ are defined in
exactly the same way as $M_{ss'}$ and $D_{\rho ss'}$ but with
$\widehat \varpi^{(s)}_{\mu\nu}$ replacing $\varpi^{(s)}_{\mu\nu}$ in
eq.~$(\ref{e42})$.

It is straightforward to show that
\begin{eqnarray}
\tau^\mu_\nu\nabla_\rho{\varpi^{(+)}}{}^\nu_\mu&=&
-\nabla_\rho\tau^\mu_\nu{\varpi^{(+)}}{}^\nu_\mu =0\>,
\nonumber \\
\nabla_\rho\tau^\mu_\nu\nabla^\rho\tau^\nu_\mu &=&
-2M^2_{++}-2D_{+\times}^2\>,
\nonumber \\
\nabla_\rho{\varpi^{(\times)}}{}^\mu_\nu\nabla^\rho\tau^\nu_\mu &\equiv&
	-2G=-4\nabla_\mu\epsilon^{(1)}_\nu\nabla^\mu\epsilon^{(2)}{}^\nu\>.
\nonumber \\
\nabla_\rho{\varpi^{(+)}}{}^\mu_\nu\nabla^\rho\tau^\nu_\mu &\equiv&
	-2H=-2\left(
	\nabla_\mu\epsilon^{(1)}_\nu\nabla^\mu\epsilon^{(1)}{}^\nu
	-
	\nabla_\mu\epsilon^{(2)}_\nu\nabla^\mu\epsilon^{(2)}{}^\nu
	\right)\>.
\label{a11}
\end{eqnarray}
Then
\begin{eqnarray}
\widehat M^2_{++} &=& M^2_{++} - \nabla_\rho\varphi \nabla^\rho\varphi
-\sin{2\varphi} H+\sin^2\varphi D_{+\times}^2\>,
\nonumber \\
\widehat M^2_{\times\times } &=& M^2_{++} - \nabla_\rho\vartheta
\nabla^\rho\vartheta -\cos^2\vartheta \nabla_\rho\varphi
\nabla^\rho\varphi
\nonumber \\
&{}&
+\cos^2\vartheta\sin2\varphi H -
\sin2\vartheta\cos\varphi G
\nonumber \\
&{}&
+\cos^2\vartheta\cos^2\varphi D^2_{+\times} +
(2\sin\varphi\nabla_\rho\vartheta
-\cos\varphi\sin{2\vartheta}\nabla_\rho\varphi) D^\rho_{+\times}\>,
\nonumber \\
\widehat M^2_{+\times } &=&
\sin\vartheta\nabla_\rho\vartheta\nabla^\rho\varphi -
\cos{2\varphi}\cos\vartheta H - \sin\vartheta\sin\varphi G
\nonumber\\
&{}&
-\left(\sin\varphi\sin\vartheta\nabla^\rho\varphi +
\cos\varphi\cos\vartheta\nabla^\rho\vartheta\right)D_{\rho+\times}
+\frac{1}{2}\sin{2\varphi}\cos\vartheta D^2_{+\times}\>.
\label{a12}
\end{eqnarray}
While $\widehat D_{\rho++}=\widehat D_{\rho\times\times}=0$ still,
\begin{equation}
\widehat D_{\rho+\times} = \sin\vartheta\cos\varphi D_{\rho+\times} +
\cos\vartheta\nabla_\rho\varphi \>.
\label{a13}
\end{equation}

In a similar fashion, we find that
\begin{equation}
\widehat I_R = -\frac{1}{2}\sum_{\vec q} \sum_{ss'}\int a^4 d^4 x
\widehat T_{ss'} \bar h_{s'}h_{s}\>,
\label{a14}
\end{equation}
with
\begin{eqnarray}
\widehat T_{++} &=& \cos2\varphi\, T_{++}\>,
\nonumber \\
\widehat T_{\times\times} &=& \left(\sin^2\vartheta -
	\cos^2\vartheta\cos2\varphi\right) T_{++}\>,
\nonumber \\
\widehat T_{+\times} &=& -\cos\vartheta\sin2\varphi \,T_{++}\>,
\label{a15}
\end{eqnarray}
where we have used
\begin{equation}
\sum_{ijkm}R^{ij}{}_{km}\varpi^{(+)}{}^m_i\tau^k_j = 0\>,\quad
\sum_{ijkm}R^{ij}{}_{km}\varpi^{(\times)}{}^m_i\tau^k_j = 0\>,\quad
\sum_{ijkm}R^{ij}{}_{km}\tau^m_i\tau^k_j = -T_{++}\>.
\label{a16}
\end{equation}

Finally, we have the gauge term,
\begin{eqnarray}
\widehat I_{g} 	&=& -\frac{1}{2}\sum_{\vec q ss'}\int a^4 d^4x
		\Bigg\{
		\widehat{\varpi}^{(s)}\widehat{\varpi}^{(s')}\nabla_\rho
		h_s\nabla^\rho \bar h_{s'}
		+
		2\widehat{\varpi}^{(s)}\nabla_\rho\widehat{\varpi}^{(s')}\bar
		h_{s'}\nabla^\rho h_s +
\nonumber \\
	&{}&
		\left(
		\nabla_\rho\widehat{\varpi}^{(s)}\nabla^\rho
		\widehat{\varpi}^{(s')}
		+
		\frac{1}{a^2}\widehat{\varpi}^{(s)}\sum_j
		\frac{d\>}{d\eta}\left(\frac{a_j'}{a_j}\right)
		\widehat\varpi^{(s')}{}^j_j
		\right)
		h_s\bar h_{s'}
		\Bigg\}\>,
\label{a17}
\end{eqnarray}
where
$\widehat{\varpi}^{(s)}\equiv\widehat{\varpi}^{(s)}{}^\mu_\mu/\sqrt2$ and
is given in eq.~$(\ref{a7})$.

Combining the three terms together, we obtain
\begin{eqnarray}
\hat I &=& \frac{1}{2}\sum_{\vec q ss'}\int a^4 d^4x \Bigg\{
		\left(
		\delta_{ss'} -
		\widehat{\varpi}^{(s)}\widehat{\varpi}^{(s')}
		\right)\nabla_\rho\bar h_{s'}\nabla^\rho h_s
\nonumber \\
&{}&
		-
		2 \widehat{\varpi}^{(s)}\nabla_\rho\widehat{\varpi}^{(s')}\bar
		h_{s'}\nabla^\rho h_s-2\widehat D_{\rho ss'} \bar
		h_{s'}\nabla^\rho h_s
		-
		\Bigg(
		\widehat M_{ss'} +\widehat T_{ss'} +
		\nabla_\rho\widehat{\varpi}^{(s)}\nabla^\rho
		\widehat{\varpi}^{(s')}
\nonumber \\
&{}&
		+
		\frac{1}{a^2}\widehat{\varpi}^{(s)}\sum_j
		\frac{d\>}{d\eta}\left(\frac{a_j'}{a_j}\Bigg)
		\widehat\varpi^{(s')}{}^j_j
		\right)h_s\bar h_{s'}\Bigg\}
\label{a18}
\end{eqnarray}
We can now see explicitly that instead of eliminating the coupling
term, this choice of gauge merely makes things worse. In fact, the
kinetic term for the two modes are now quite different than what we
would expect and additional coupling terms now appear. It is, however,
important to note that these additional terms have the same origin as
those obtained by using the
previous gauge. Namely, they all come from linear combinations of
eqs.~$(\ref{e51c})$-$(\ref{51e})$ and arise from the fact that
the directions of the polarization vectors are continually changing
with time.

\end{appendix}


\begin{references}
\bibitem[7]{com1} This choice of parametrization is different from
		  that found in $\cite{BLK}$ but we shall find it
		  more convenient to use.
\bibitem[8]{com2} For example, we could not have taken
		  $\sqrt{T_{\mu\nu} T^{\mu\nu}}$ since the square of
		  $T_{\mu\nu}$ can vanish even though $T_{\mu\nu}$
		  does not.
\end{references}
\end{document}